\documentclass{iaus}
\usepackage{graphicx}

\title[Substructure in CDM Haloes]{Predicting Substructure in CDM Haloes}

\author[Taylor \& Babul]%
{James E. Taylor$^1$%
  \thanks{PPARC Fellow},
Arif Babul$^2$}

\affiliation{$^1$Denys Wilkinson Building, 1 Keble Road, Oxford OX1 3RH, United Kingdom \\email: jet@astro.ox.ac.uk\\[\affilskip]
$^2$Elliott Building, 3800 Finnerty Road, Victoria, BC, V8P 1A1, Canada \\email: babul@uvic.ca}

\pubyear{2004}
\volume{225}
\pagerange{1--8}
\date{?? and in revised form ??}
\jname{Impact of Gravitational Lensing on Cosmology}
\editors{Mellier, Y. \& Meylan,G. eds.}
\begin{document}

\maketitle

\begin{abstract}
Observations of multiple-image gravitational lens systems 
suggest that the projected mass distributions of galaxy haloes 
may contain substantial inhomogeneities. The fraction of the halo
mass in dense substructure is still highly uncertain, but could
be as large as a few percent. While halo substructure is seen
in numerical simulations of CDM haloes, little of this substructure 
survives in the innermost regions of haloes, and thus the 
observational claims for substructure at small projected radii 
are slightly surprising. There is evidence, however, that even 
the highest-resolution simulations published to-date are still 
limited by numerical effects that heat and disrupt substructure
artificially in high-density regions. By comparing numerical
and semi-analytic (SA) models of halo substructure, we show that
current simulations probably underestimate the mass fraction 
in substructure at small projected radii, by a factor of at least
2--3. We discuss the prospects for using lensing observations
as a fundamental test of the nature of dark matter.
\end{abstract}

\firstsection % if your document starts with a section,
\section{Introduction: The case for substructure in observed haloes}

There is now overwhelming evidence that the matter content of the 
Universe is in large part non-baryonic (Spergel et al.\ 2003). The
strongest contender for this extra component is some sort of weakly 
interacting cold dark matter (CDM), possibly in the form of a relic 
supersymmetric particle. One of the most surprising properties of CDM
is its tendency to cluster gravitationally on subgalactic 
(and possibly even possibly subsolar) scales. Detecting subgalactic
lumps of dark matter would be a dramatic confirmation of CDM theory, 
and should be considered a top priority for observational astrophysics.

Strong lensing has long been recognised as a method for identifying
dark,  compact objects on subgalactic scales (e.g. Press \& Gunn 1973;
Blandford \& Jaroszynski 1981; Wambsganss \& Paczynski 1992). Interest
in these tests was recently revived by Mao \& Schneider (1998), who
proposed that some form of substructure -- either a normal baryonic
feature such as a spiral arm or a globular cluster, or substructure
within the dark matter halo of the system -- might explain why the
flux ratios of the different components of the multiple-image system
B1422+231 disagreed so strongly with the predictions of models
assuming smooth lens potentials.

The idea that anomalous flux ratios might be due to substructure was
reexamined more recently by several authors (Metcalf \& Madau 2001;
Chiba 2002; Dalal \& Kochanek 2002). Since these initial papers, there
has been extensive work trying to extract more quantitative
information  from the observations (e.g.\ Metcalf \& Zhao 2002;
Metcalf 2002;  Bradac et al.\ 2002; Schechter \& Wambsganss 2002;
Chen, Kravtsov \& Keeton 2003; Keeton, Gaudi \& Peters 2003),  but
there is increasing concern that other phenomena such as microlensing
may be responsible for the deviations observed (Schechter \&
Wambsganss 2002),  or that they may simply be due to insufficiently
general modelling  of the lens potential (Evans \& Witt 2003). Currently
there still seems good  evidence for genuine anomalies in a few
systems (Moustakas \& Metcalf 2003;  Metcalf et al. 2004; Metcalf these
proceedings), suggesting substructure in the projected mass
distribution at the level of a few percent, but  continued
observations at many different wavelengths will be required to
disentangle the effects of substructure from microlensing,
scintillation  or other phenomena. The observational effort is
worthwhile, however, as it  may provide the first hard evidence to
justify  one of the main assumptions of our current cosmological model,
the cold, collisionless nature of dark matter.

\section{A new approach to modelling CDM haloes}

Whatever the status of the substructure problem observationally, it is not 
clear that there is a robust theoretical prediction with which to compare the 
lensing results. Dark matter haloes form through the gravitational collapse of 
diffuse dark matter, as well as the hierarchical merging of smaller haloes. 
The process is sufficiently non-linear that most of our understanding of it 
comes from numerical simulations. The strong lensing anomalies depend on 
the net mass fraction in relatively low-mass substructure 
($10^5 M_{\odot}$--$10^7 M_{\odot}$), projected on the central few 
kiloparsecs of galaxy haloes. This is close to, or beyond, the formal 
resolution 
limit of most current simulations, and even in those simulations that 
can resolve structures
on this scale, serious doubts remain as to the completeness of the results in 
the innermost parts of the halo. Thus, while simulations currently predict 
less central substructure than inferred from observations, this may 
be partly due to their limited resolution.

To study halo substructure on smaller scales or very close to the centre
of the halo, we have developed an alternative, semi-analytic model 
(Taylor \& Babul 2004; Taylor \& Babul in preparation). This model includes 
several distinct components. First, merger histories for a large number
of individual haloes are generated randomly, using Press-Schechter statistics
and the merger-tree algorithm of Somerville and Kolatt (1999), together with
a correction for higher-order substructure developed in 
Taylor \& Babul (2004). 
Each merging subhalo is then placed on a random orbit starting at the virial 
radius of the main system, and evolved using the analytic model of satellite 
dynamics described in Taylor \& Babul (2001), experiencing orbital decay due 
to dynamical friction, and heating and stripping due to tidal forces. The 
properties of the main system change dynamically over time, and the 
formation of a galaxy can also be modelled schematically, as described in 
Taylor \& Babul (2003), further modifying the central mass distribution.

Overall, this model provides a computationally efficient way of simulating
the hierarchical assembly of galaxy or cluster haloes, and the evolution
of their substructure. Because it performs only a few calculations per
lump of dark matter (or `subhalo') merging with the main system 
it can be used to track the evolution of many thousands of subhaloes
in a typical system, providing complete information about halo substructure
down to masses around $10^5$--$10^6 M_{\odot}$. 

\section{Results: The outer halo}

To test the accuracy of the SA model, we have used it to 
generate a large set of galaxy haloes and compared their substructure
to the substructure found in a set of high-resolution simulations of
halo formation by Ghigna et al.\ (1998, 2000) and 
Moore et al.\ (1999a, 1999b). These include the galaxy-mass haloes
`Andromeda' and the `Milky Way' (the `Local Group' -- Moore et al.\ 1999b)
and cluster-mass haloes  `Coma', `Virgo I' and `Virgo II' (Virgo IIa and b
are two different outputs from the same simulation). Each was resolved
with $\sim\,10^6$ particles of more, and with a softening length of
less than 1\% of the virial radius. 

\begin{figure}
\centerline{
\scalebox{0.38}{
 \includegraphics{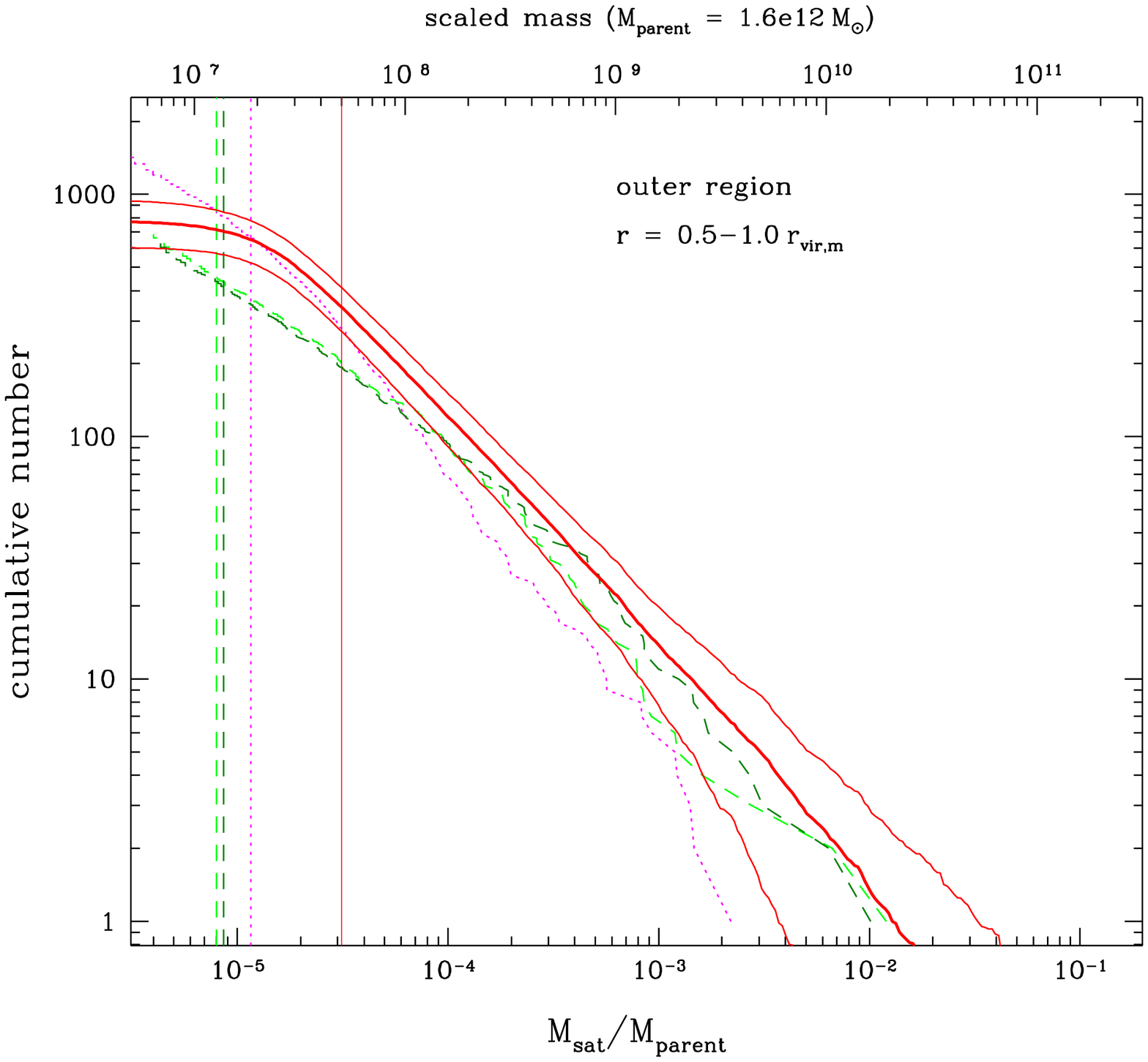}
}
\scalebox{0.35}{
 \includegraphics{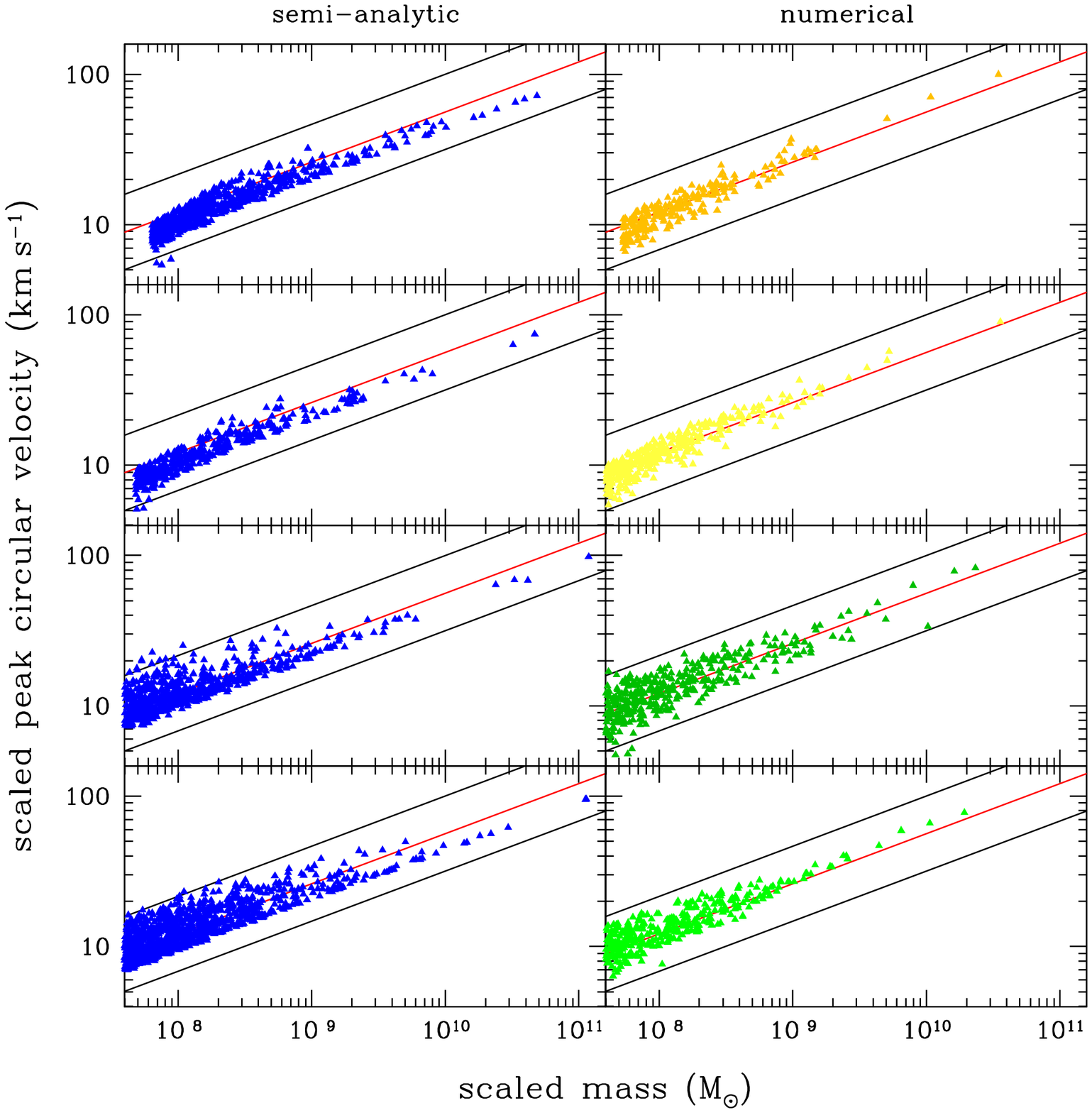}
}
}
  \caption{(Left) The cumulative mass function predicted by the
SA model in the outer parts of the halo. The thick lines 
show the average result for a hundred SCDM merger trees at $z = 0$. The
thin solid lines show the 1-$\sigma$ variance for this set. The dashed 
and dotted lines are the normalised cumulative mass functions measured 
in three high-resolution simulations (dashed lines -- Virgo IIa and
IIb; dotted lines -- Coma). The vertical lines indicate the resolution
limit of the SA trees (solid) and the 32-particle mass limits of the 
simulations (dotted and dashed). 
(Right) The distribution of subhaloes as a function of their mass
and of their peak velocity, in the SA model (left-hand plots) and the 
simulations (right-hand plots; values are scaled to the SA halo mass 
and velocity).}\label{fig:1}
\end{figure}

The left-hand plot in figure~\ref{fig:1} compares the cumulative 
distribution of subhalo
masses in the outer regions of the SA haloes (thick solid line) with
the simulations (dashed or dotted lines). The masses have been scaled
to the mass of the parent halo in the SA model, $1.6 \times 10^{12} M_{\odot}$,
for comparison.  Overall we find an excellent match in the
outer regions of the halo. The simulations have an average amplitude 
about 20\% lower than the SA average, but
this is only 1--2 times the halo-to-halo scatter (thin solid lines).  We
note that this agreement is achieved  without adjusting any free
parameters -- the parameters in the  semi-analytic model have all been
fixed previously by other considerations,  as discussed in paper
Taylor \& Babul (2004).

The right-hand plot in figure~\ref{fig:1} shows that the internal 
structure of subhaloes is
also very similar. The peak circular velocity of each system is plotted
versus its mass, for the SA models (left-hand plots) and the numerical
models (right-hand plots). Values are scaled as in figure~\ref{fig:1}.
The effects of softening and shot noise in the number of particles within 
the peak radius of each subhalo have been added to SA results, since
these effects are present in the numerical data. With this correction, 
we find very good agreement between the distributions of subhalo
properties.
 
\section{Results: The inner halo}

\begin{figure}
\centerline{
\scalebox{0.4}{
 \includegraphics{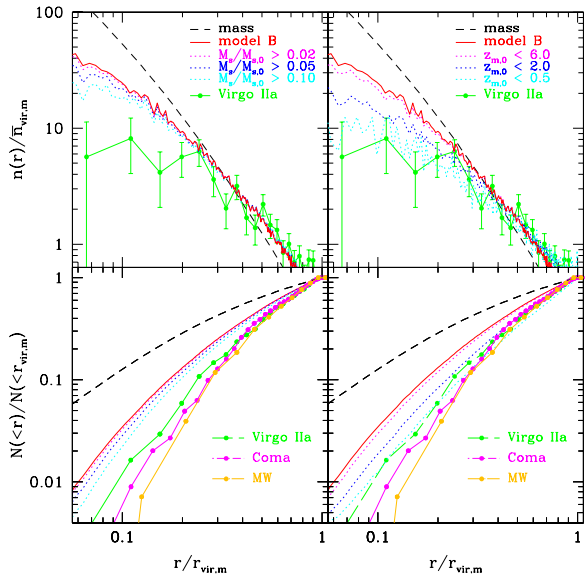}
}
}
  \caption{ Top panels: The number density of subhaloes in the SA haloes
(upper solid lines) and in three simulations (connected points with error 
bars). To avoid incompleteness, the results are cut at the equivalent of 
$5 \times 10^{7} M_{\odot}$. In each case the density is relative 
to the mean within the virial radius. The dashed line shows the density 
profile of the main halo, normalised to the mean within the virial
radius. 
Bottom panels: The cumulative number of subhaloes vs.\ radius,
normalised to the number within the virial radius, for the same mass
cuts as in the top panel. The dashed lines show the mass of the main
halo interior to a given radius, normalised to the mass within the
virial radius. The dotted lines in the left-hand panels show the
results of ignoring highly stripped systems; the dotted lines in the 
right-hand panels show the results of ignoring old systems.}\label{fig:2}
\end{figure}

In the inner halo, on the other hand, the SA model run with the same
parameters as above predicts 2--3 times more substructure than is seen
in the numerical simulations. Moreover, the distribution of substructure
is much more centrally concentrated than in the simulations. 
Figure~\ref{fig:2} compares the radial distribution of subhaloes
in the SA models (solid and dotted lines) and three simulations
(dashed or dot-dashed lines + data points). The smooth dashed
curve shows the overall mass distribution of the parent halo.
While both SA and numerical subhaloes are antibiased with respect
to the background mass distribution, the SA model predicts a central 
density of substructure 3--4 times higher than the numerical value.
Removing highly-stripped systems (left-hand panel) does not affect
this result very strongly, but removing old systems (right-hand panel)
does. Thus the SA model predicts the existence of an extra population
of old systems, deep in the centre of the halo.

\begin{figure}
\centerline{
\scalebox{0.4}{
 \includegraphics{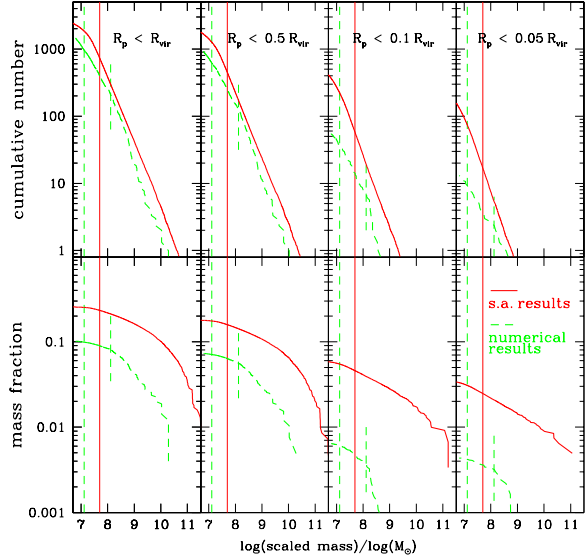}
}
}
  \caption{
(Top panel) Cumulative mass functions for subhaloes
within some projected radius $R_{\rm p}$, for the SA model (solid lines)
and Virgo IIa (dashed lines). (Bottom Panel) The fraction of the
projected mass within $R_{\rm p}$ contained in subhaloes of mass $M$
or larger. The numerical results are the average over three different
projections. Vertical lines indicate the resolution limit of the merger
tree (solid) and the 32-particle mass limits of the simulation (dashed).}\label{fig:3}
\end{figure}

The importance of these results for lensing detections of halo
substructure is illustrated in figure~\ref{fig:3}. This compares
the projected mass function within various radii around a galaxy halo
(top panels), and the fraction of the projected surface density in
substructure (bottom panels), for the SA model and the highest-resolution
simulation (solid and dashed lines respectively). The offset of $\sim\,2$ 
between the 
two, when averaged over large projected radii (leftmost panel), grows to 
roughly an order of magnitude at small projected radii (3rd and 4th panels).

\section{Summary and future prospects}%\label{sec:concl}

When observations can reliably determine the clustering of dark matter
on subgalactic scales, they will shed light on a number of important
questions in fundamental physics. The nature and amplitude of 
small-scale CDM structure depends on the spectrum of perturbations generated 
towards the end of inflation, which in turn depends on the primordial spectral 
index, the shape of inflationary potential, and the physics of reheating. 
The subsequent growth of these perturbations also depends on the equation of 
state of the universe at the quark-hadron transition and at nucleosynthesis. 
Finally, the evolution of small-scale structure at late times is a sensitive 
test of dark matter physics, such as interactions or annihilation. These 
small-scale properties will also have an important effect on the growth of 
visible 
structure, especially at very high redshift during the epoch of reionization.

To tap the potential of recently developed methods for detecting
substructure in strongly lensed systems, we need robust predictions for
the behaviour of dark matter on very small scales, at very high densities,
over cosmological timescales. Achieving this goal remains a challenge
for current numerical simulations of structure formation.  
We have presented initial results from a semi-analytic model which
uses halo merger trees and satellite dynamics to model the
properties of substructure within dark matter haloes. Without any
adjustment of free parameters, this model matches the results of
high-resolution simulations very closely in the outer parts of haloes,
where the simulations are most likely to be accurate. In the inner parts,
however, it predicts central densities of substructure 3--4 times 
higher than those found in simulations. This may help explain
the very high levels of substructure inferred from recent lensing
observations. 

\begin{acknowledgments}
The authors wish to thank E. Hayashi, S. Ghigna, B. Moore, J. Navarro 
and T. Quinn for providing the substructure data from their simulations. 
JET gratefully acknowledges support from PPARC UK. AB gratefully 
acknowledges support from NSERC Canada, through the Discovery and the 
Collaborative Research Opportunities (CRO) programs.
\end{acknowledgments}

\begin{discussion}

\discuss{L. King}{If you take your predicted substructure into account,
will it explain the observations?}

\discuss{J. Taylor}{Yes, or at least it almost will. The status of the
observations  is a bit unclear, but in the few systems were we do have
strong evidence for  substructure, the inferred level is high --
within a few percent of the virial radius, a few percent of the
projected mass density appears to be in substructure. This is at
least ten times more than the simulations predict, but would be close
to or just above the level the SA model predicts. The SA model also
predicts a factor of 3 variation from one system to the next,
however, so we really need to observe more systems.}
\end{discussion}

\end{document}